\documentclass[a4paper,conference]{IEEEtran}
\ifCLASSINFOpdf
\else
\fi
\usepackage[pdftex]{graphicx}
\usepackage{cite}
\usepackage[caption=false,font=footnotesize]{subfig}

\newcommand{\wfig}[1]{Fig.\ref{fig:#1}}
\newcommand{\wfigure}[1]{Figure \ref{fig:#1}}

\begin{document}
%
\title{Performance Evaluation of Two-layer lossless HDR Coding using Histogram Packing Technique under Various Tone-mapping Operators}

\author{\IEEEauthorblockN{Hiroyuki KOBAYASHI}
\IEEEauthorblockA{
 Tokyo Metropolitan College  of Industrial Technology,\\
 Email: hkob@metro-cit.ac.jp}
\and
\IEEEauthorblockN{Hitoshi KIYA}
\IEEEauthorblockA{
 Tokyo Metropolitan University\\
Email: kiya@tmu.ac.jp}
}


%


\maketitle



%
\IEEEpeerreviewmaketitle

\section*{Abstract}

We proposed a lossless two-layer HDR coding method using a histogram packing technique.
The proposed method was demonstrated to outperform the normative JPEG XT encoder, under the use of the default tone-mapping operator.
However, the performance under various tone-mapping operators has not been discussed.
In this paper, we aim to compare the performance of the proposed method with that of the JPEG XT encoder under the use of various tone-mapping operators to clearly show the characteristic difference between them.

\begin{IEEEkeywords}
JPEG XT, HDR coding, Reversible coding, Tone-mapping operator
\end{IEEEkeywords}

\section{Introduction}
Compression methods for HDR (High Dynamic Range) images are expected to meet the rapid growth in HDR image applications.
Generally, HDR images have a long bit depth of pixel values and wide color gamut\cite{ReinhardBook,8026195,ArtusiBook01,BADC11}.
These characteristics are suitable for various applications, such for medicine and art.
HDR images are often required to be reversely encoded in the applications.
They should be compressed with almost no coding loss.

ISO/IEC IS 18477-8:2016\cite{JPEGXTpart8}, namely JPEG XT Part 8, provides how to decode losslessly or near-losslessly compressed HDR images.
The JPEG XT is a two-layer coding method and is backward compatible with the legacy JPEG standard\cite{JPEG-1}, and some extended coding methods have been studied\cite{7991151,Kobayashi2019ICCEASIAPP}.
Our research group proposed a novel two layer lossless HDR coding method using a histogram packing technique, which has the backward compatibility to JPEG, referred to as HP coder\cite{8351220,IEICETwoLayerLosslessHDR}.
The packing technique allows us to improve the performance of lossless compression for HDR images when images have the histogram sparseness\cite{Minewaki2017a,KIYAJan2018}.
The HP coder was demonstrated to outperform the normative JPEG XT encoder.
However, the coding performance under various Tone-mapping operators (TMOs) has not been discussed.

In this paper, we aim to compare the performance of the novel method with that of the JPEG XT encoder under the use of various tome-mapping operators.
It is shown that the HP coder has not only a higher compression performance than the JPEG XT one even when various TMOs are used, but also the compression performance is not affected by a kind of TMOs.

\section{HDR coding under the various TMOs}

\subsection{Overview of two-layer lossless HDR coding}
The coding procedures of HP coder and JPEG XT are illustrated in \wfig{proposed}, respectively.
HP coder has the same coding-path to generate the base layer, which is used to produce LDR (Low Dynamic Range) images having a backward compatibility with the legacy JPEG standard, as that of JPEG XT except that the refinement scan is not carried out.
For the residual layer, which has the residual data generated by subtracting the partially decoded base layer from the original HDR image, the histogram of each color component is packed after the color space conversion by a histogram packing technique in HP code.
The packed residual data is then compressed using an arbitrary lossless image encoder, such as JPEG 2000 or JPEG XR.

\subsection{Tone-mapping operation with a TMO}

\begin{figure*}[t]
 \centering
 \includegraphics[width=0.7\textwidth]{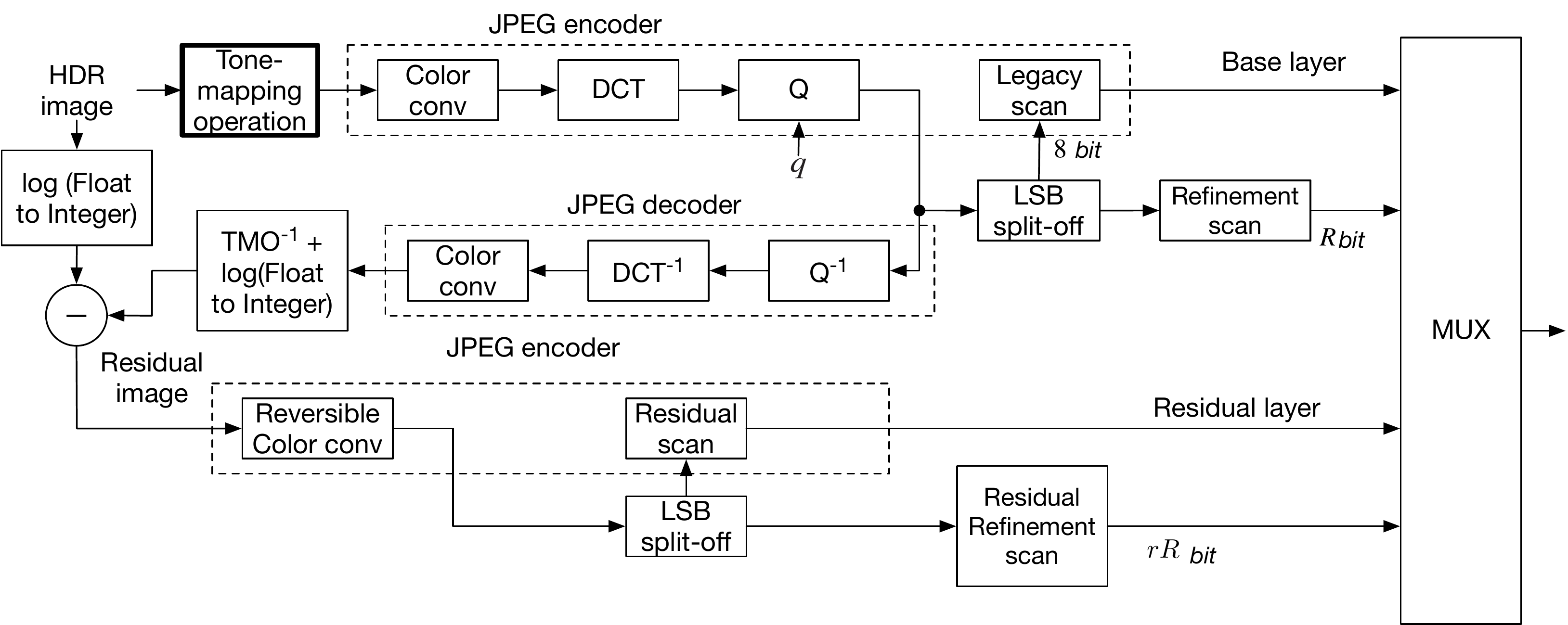} \\
 (a) XT coder\\
 \includegraphics[width=0.7\textwidth]{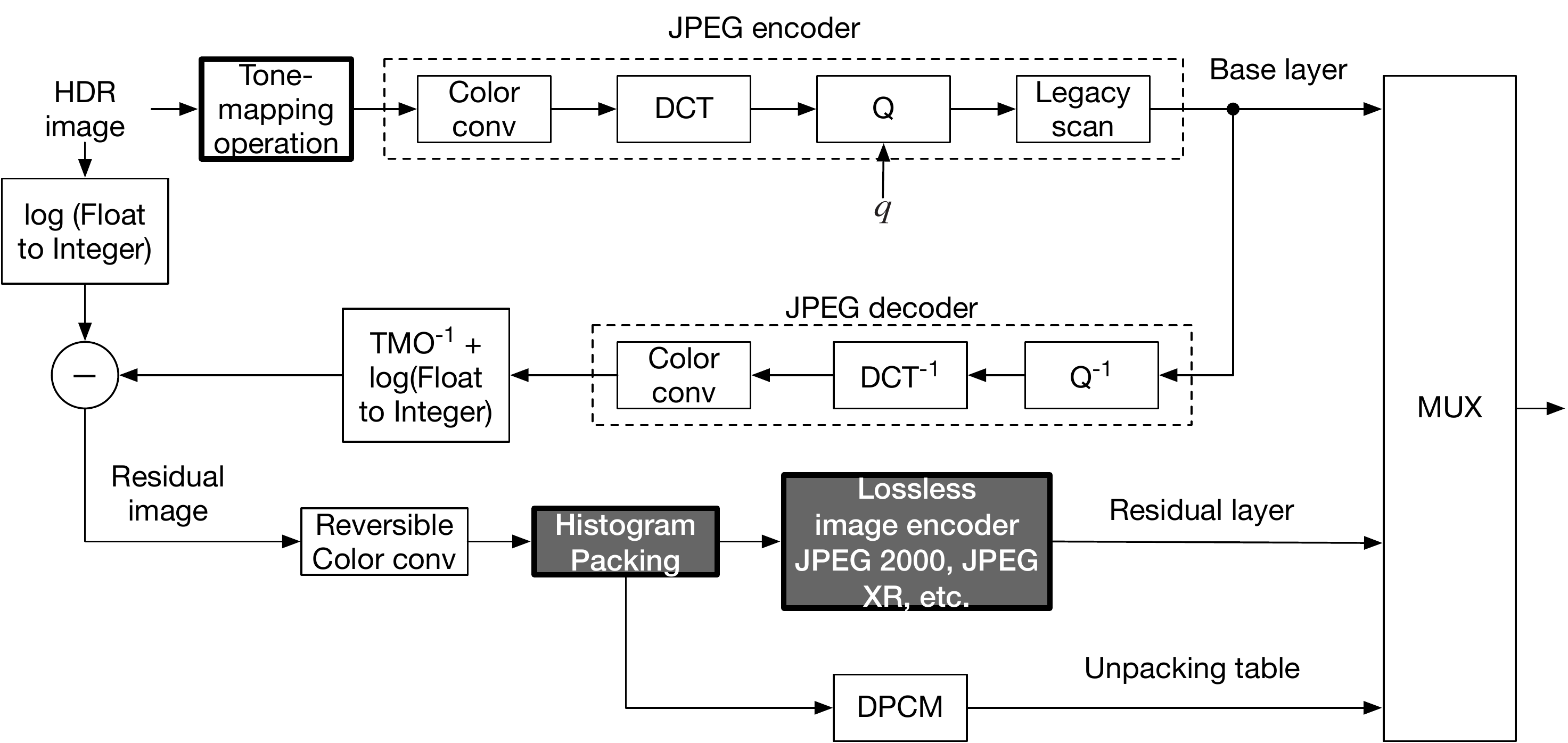} \\
 (b) HP coder\\
 \caption{Block diagram of XT coder and HP one}
 \label{fig:proposed}
\end{figure*}

Both JPEG XT and HP coders can be applied to any TMO for `Tone-mapping operation' in \wfig{proposed}.
The selection of TMOs affects not only the quality of LDR images, i.e. the base layer, but also the compression performance of HDR images.
In this paper, we aim to evaluate the influence of the TMO selection.

\section{Experimental results}
To evaluate the influence of the TMO selection, we compared the image quality of LDR images and the coding performance of HDR images under the use of four
TMOs: JPEG XT default, Reinhard (Global), Reinhard (Local), and Drago TMO.
For JPEG XT, the reference software\cite{JPEGXTpart5} provided by the JPEG committee was used.
For HP coder, a software was prepared by modifying the reference software, where Kakadu software\cite{Kakadu} (JPEG 2000 codec) was also used for the residual path as the lossless image encoder in HP coder.

\subsection{Comparison of LDR image quality}

\begin{figure}[t]
 \centering
 \includegraphics[width=0.5\textwidth]{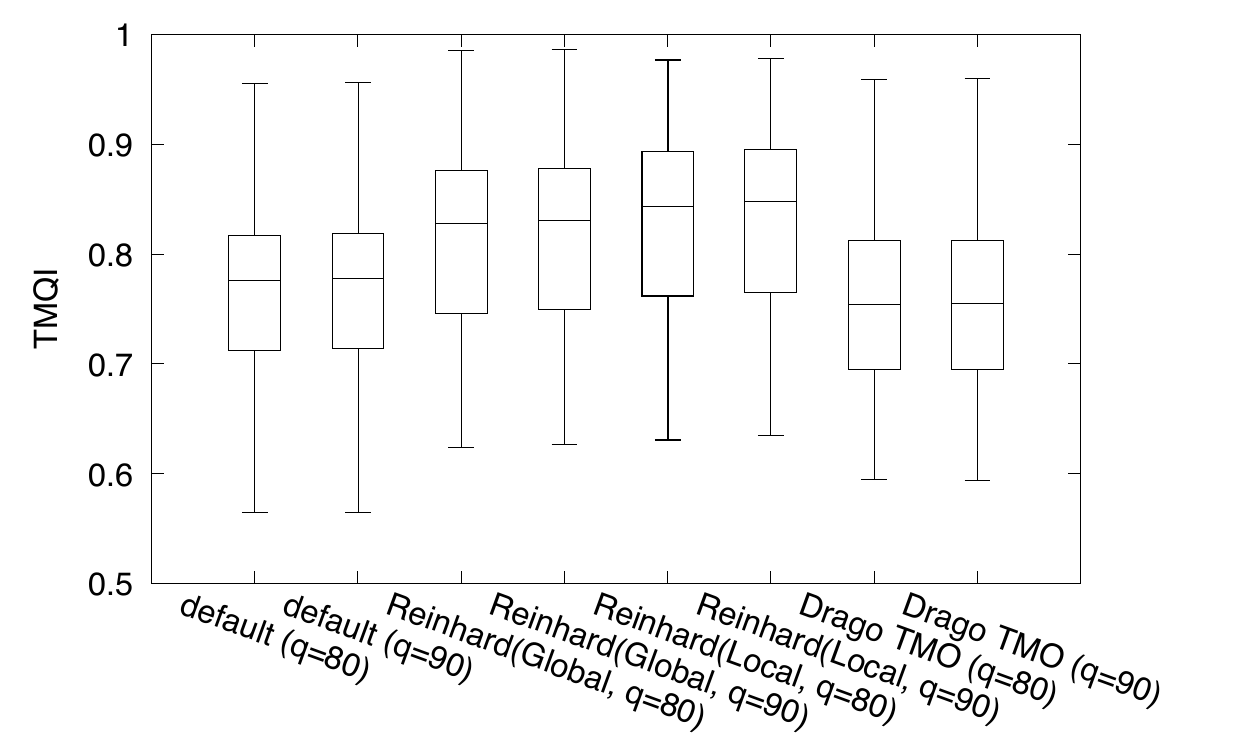}\\
 \caption{Boxplot of TMQI of proposed method.
  Box indicates three quartiles of data, and whiskers indicates lowest datum within 1.5 interquartile range (IQR) of lower quartile, and highest datum still within 1.5 IQR of upper quartile.}
 \label{fig:compTMQI}
\end{figure}

\wfigure{compTMQI} shows boxplots of the quality of LDR images under the use of HP coder in terms of Tone Mapped Image Quality Index (TMQI),which is a well known objective quality assessment algorithm for tone mapped images\cite{5651778}.
The boxplot calculated using 105 HDR images collected from the Fairchild HDR image survey\cite{fairchild2007hdr}.
From the \wfig{compTMQI}, the quality of LDR images was affected by the selection of TMO, and the default TMO did not give the best quality in various TMOs.

\subsection{Comparison of compression performance}
\begin{figure}[t]
 \centering
 \includegraphics[width=0.5\textwidth]{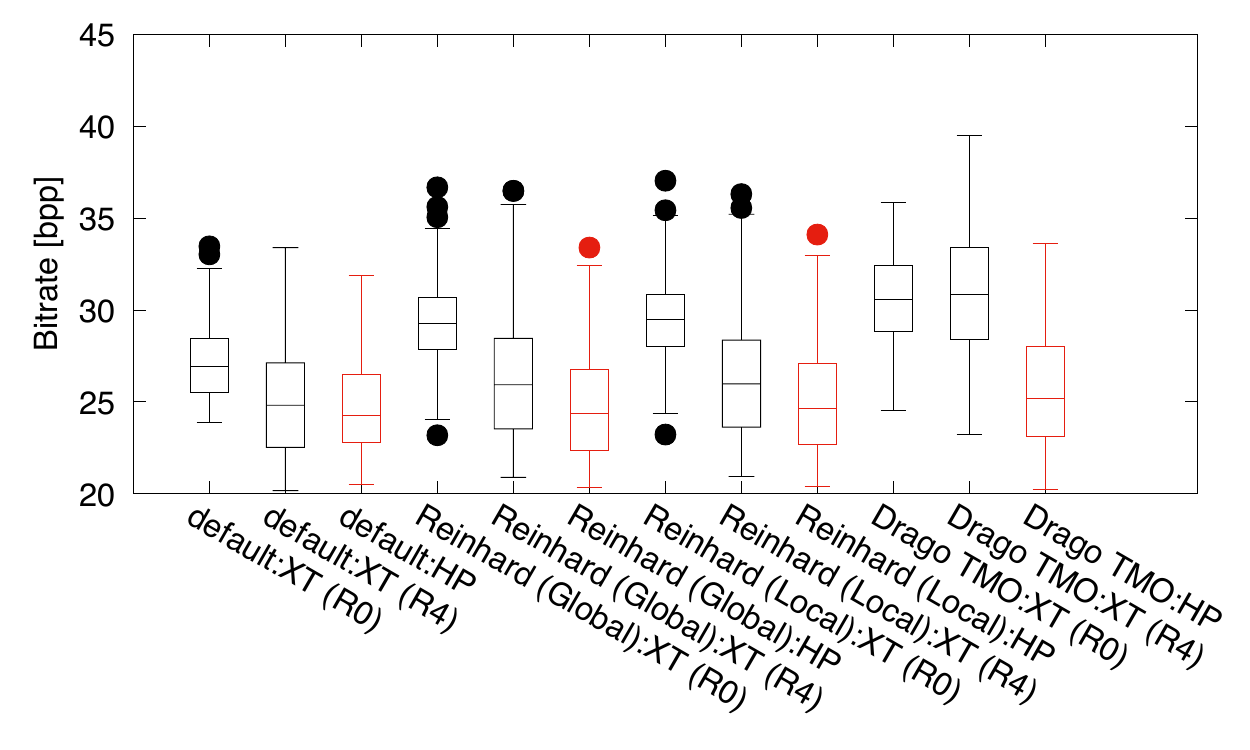} \\
 (a) $q= 80$\\
 \includegraphics[width=0.5\textwidth]{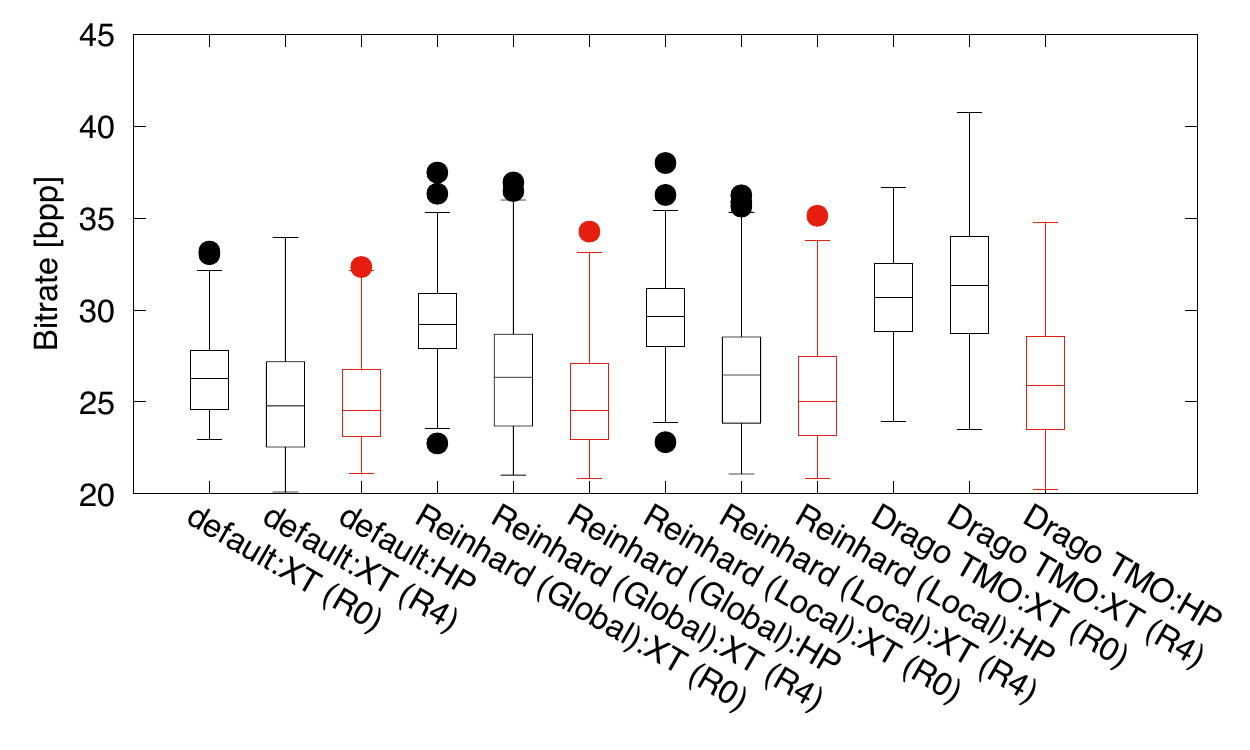} \\
 (b) $q=90$\\ 
 \caption{Boxplot of bitrates of HP coder and XT coder ($R=0$ or $R=4, rR=0$).
 Data not included between whiskers should be plotted as outlier with dot.}
 \label{fig:compBitrate}
\end{figure}

\wfigure{compBitrate} shows boxplots of total bitrates calculated using the same 105 HDR images.
We chose $R = 0$ or $4$, $rR = 0$ and $q = 80$ or $90$ for JPEG XT, and we chose only $q = 80$ or $90$ for the HP one, where $R$, $rR$ and $q$ are the number of bits used for the refinement scan, the number of bits used for the residual refinement scan, and the quantization parameter in the base layer, respectively\cite{IEICETwoLayerLosslessHDR}.
From \wfig{compBitrate}, HP coder outperformed JPEG XT under all TMOs.
Moreover, the compression performance of JPEG XT was affected by the selection of TMOs.
In contrast, HP coder was almost no effected by the selection.

\section{Conclusions}
We discussed the performances of two two-layer lossless HDR coding methods under various tone-mapping operations: JPEG XT and HP coder.
The experimental results demonstrated that HP coder outperformed JPEG XT under all TMOs in terms of the lossless compression performance.
In addition, the default TMO was also demonstrated not to give the best quality of LDR images in various TMOs.

\bibliographystyle{IEEEbib}

\bibliography{library}

\end{document}